\documentclass[amsmath,amssymb,prc,final,showpacs,twocolumn]{revtex4-1}
 
\usepackage{bm,hyphenat,xspace}
\usepackage{graphicx,epsfig}

\newcommand {\mcu}{\mathcal{U}}

\newcommand{\cm}{\mathrm{c\!\:\!.m\!\:\!.}}
\newcommand{\He}{{}^3\mathrm{He}}
\newcommand{\Hh}{{}^3\mathrm{H}}

\newcommand{\Hdp}{${}^2\mathrm{H}(d,p){}^3\mathrm{H}$\,\,{}}
\newcommand{\Hdn}{${}^2\mathrm{H}(d,n){}^3\mathrm{He}$\,\,{}}

\begin{document}

\title {Four-body calculation of ${}^2\mathrm{H}(d,p){}^3\mathrm{H}$ and 
${}^2\mathrm{H}(d,n){}^3\mathrm{He}$ reactions above breakup threshold}
  
\author{A.~Deltuva} 
\email{arnoldas.deltuva@tfai.vu.lt}
\affiliation
{Institute of Theoretical Physics and Astronomy, 
Vilnius University, Saul\.etekio al. 3, LT-10222 Vilnius, Lithuania
}

\author{A.~C.~Fonseca} 
\affiliation{Centro de F\'{\i}sica Nuclear da Universidade de Lisboa, 
P-1649-003 Lisboa, Portugal }

\received{\today} 

\begin{abstract}
Nucleon transfer reactions in deuteron-deuteron 
collisions at energies above the three- and four-body breakup threshold
are described using exact four-body equations for transition operators
that are solved in the momentum-space framework.
Differential cross sections, analyzing
powers, polarizations, and spin transfer coefficients are 
obtained using realistic two-nucleon potentials
and including the Coulomb repulsion between protons.  
Overall good agreement between predictions and experimental data 
is found.  Most remarkable discrepancies are seen around the minima of
 the differential cross section at higher energies and in the outgoing
nucleon polarization at lower energies.  

\end{abstract}

 \maketitle

\section{Introduction \label{sec:intro}}

In the last 10 years significant progress has taken place
in exact ab initio calculations of two-cluster scattering
in the four-nucleon system, both below
\cite{viviani:01a,lazauskas:04a,deltuva:07a,deltuva:07b,deltuva:07c,kievsky:08a,lazauskas:09a,viviani:10a,viviani:13a}
and above 
\cite{deltuva:12c,deltuva:13c,deltuva:14a,deltuva:14b,deltuva:15a,lazauskas:15a,deltuva:15c,deltuva:15e} 
the breakup threshold.
In the case of multichannel reactions, i.e., processes with open rearrangements and/or breakup channels,
the most advanced calculations were performed by solving the
Alt, Grassberger, and Sandhas (AGS) equations \cite{grassberger:67,alt:jinr} for the 
four-particle transition operators in the  momentum-space framework.
Realistic nucleon-nucleon ($NN$) force models were included as well as the
Coulomb interaction between protons, and the effective three- and
four-nucleon ($3N$ and $4N$) forces through 
explicit $\Delta$-isobar excitation.
Comparing the predictions with the available experimental data one could say
that the realistic force models  provide a satisfactory explanation
of the cross sections and spin observables over a broad energy regime
up to about 30 MeV, reproducing the proper energy dependence
and  changes in shape of the observables as the energy
rises. Agreement with data can be in some cases paradigmatic, given the
complex structure of some observables that display several maxima and
minima, a feature that is not so often observed in three-nucleon
data.  However,  there are also a few 
disagreements with the data whose magnitude is sensitive to the 
force model used and may serve as a fertile ground for the investigation 
of nuclear interactions.  Still missing in this study is the extension of the nucleon transfer reactions
${}^2\mathrm{H}(d,p){}^3\mathrm{H}$ and  ${}^2\mathrm{H}(d,n){}^3\mathrm{He}$ initiated
by deuteron-deuteron collisions at  energies well above the breakup threshold 
 where the existing differential cross 
section and analyzing power data
\cite{gruebler:72a,gruebler:81a,brolley:57,vanoers:63b,thornton:69,drosg:15,dietrich:72,%
guss:83,konig:79,hardekopf:72b,spalek:72,salzman:74,gruebler:74}
 show a complicated structure  
involving several maxima and minima that change with the energy. 
Therefore in the present manuscript we aim to investigate
${}^2\mathrm{H}(d,p){}^3\mathrm{H}$ and  ${}^2\mathrm{H}(d,n){}^3\mathrm{He}$
reactions up to about 25 MeV deuteron beam energy.

In Sec. II we present the theoretical framework.
  Differential cross section and spin observable
 results are reported in Sec. III and a summary is presented in Sec. IV.

\section{Theory \label{sec:eq}}

\begin{figure*}[t]
\begin{center}
\includegraphics[scale=0.68]{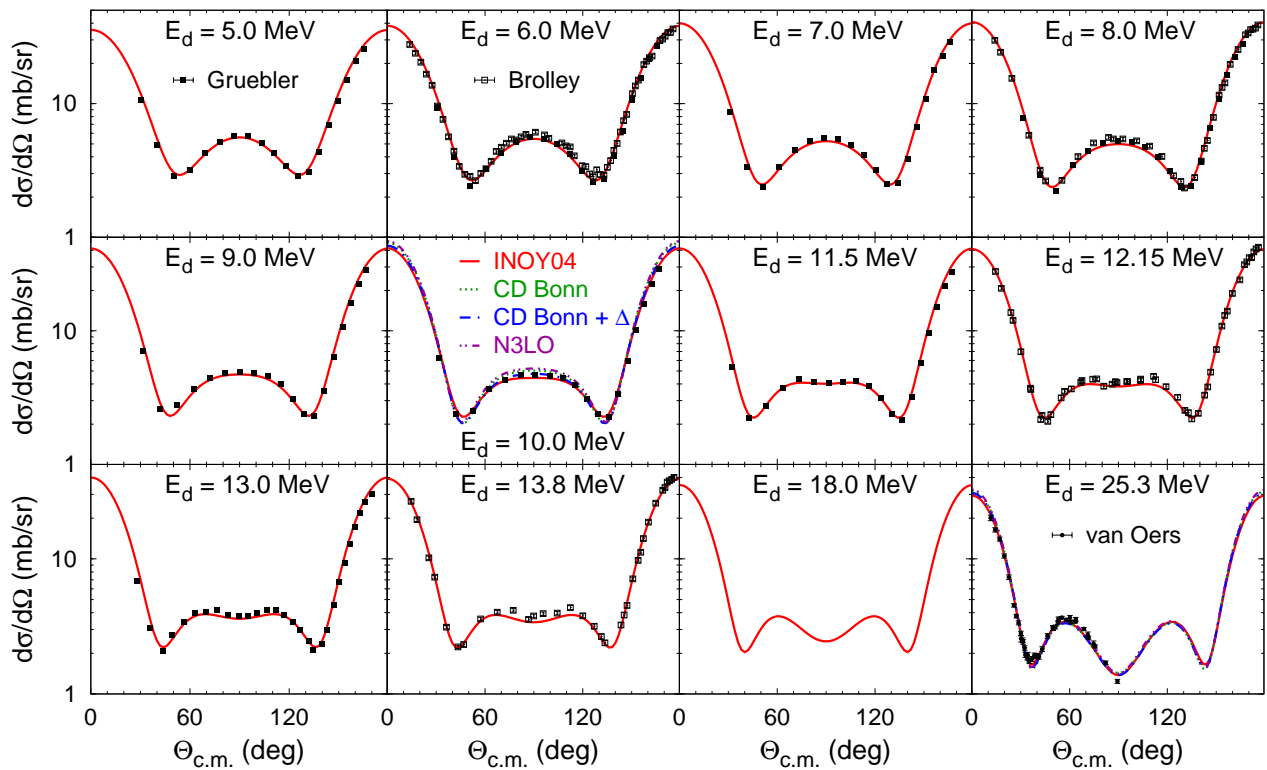}
\end{center}
\caption{\label{fig:sp} (Color online)
Differential cross section for the \Hdp reaction as a function
of the c.m. scattering angle at 
deuteron beam energies ranging from 5.0 to 25.3 MeV.
Results are obtained using the INOY04 potential (solid curves), and,
at 10.0 and 25.3 MeV, also the CD Bonn + $\Delta$ (dashed-dotted 
curves), CD Bonn (dotted curves), and N3LO (double-dotted-dashed curves)
potentials. Experimental data  are from 
Refs.~\cite{gruebler:72a,gruebler:81a} ($\blacksquare$),  
\cite{brolley:57} ($\Box$),
and \cite{vanoers:63b} ($\bullet$). }
\end{figure*}

The four-particle collision process is described  by the exact 
AGS equations \cite{alt:jinr,grassberger:67} 
for the transition operators $\mcu_{\beta\alpha}$ whose
components are labeled according to the chains of partitions. 
Given that in the isospin formalism neutrons and protons are  
treated as identical particles,
there are only two chains of partitions  that
can be distinguished by the
two-cluster partitions, one ($\alpha =1$) being of the $3+1$ type, 
i.e., (12,3)4, and
another ($\alpha =2$) being of the $2+2$ type, i.e., (12)(34).
For  nucleon-trinucleon scattering in previous works
we solved  the symmetrized AGS equations for $\mcu_{\beta 1}$, but for  
reactions initiated by two deuterons, transition operators $\mcu_{\beta 2}$ are required. 
In both cases the AGS equations share the same kernel but differ in the driving term.
Thus, in the present work we solve the integral equations
\begin{subequations} \label{eq:U}
\begin{align}  
\mcu_{12}  = {}&  (G_0  t  G_0)^{-1}  
 - P_{34}  U_1 G_0  t G_0  \mcu_{12} + U_2 G_0  t G_0  \mcu_{22} , 
\label{eq:U12} \\  
\mcu_{22}  = {}& (1 - P_{34}) U_1 G_0  t  G_0  \mcu_{12} . \label{eq:U22}
\end{align}
\end{subequations}
Here $t$ is the two-nucleon transition matrix, $U_1$ and $U_2$ are the
transition operators for  the 1+3 and 2+2 subsystems,
$P_{34}$ is the permutation operator of particles 3 and 4, and
$G_0 = (E+i\varepsilon-H_0)^{-1}$ is the free 
four-particle resolvent at the available energy $E$, whereas $H_0$ is the
free Hamiltonian.
Although the physical scattering process corresponds to $\varepsilon \to +0$, 
the complex energy method uses a finite  $\varepsilon$ value when solving
the AGS equations numerically. The physical scattering amplitudes are then obtained 
by extrapolating finite  $\varepsilon$ results to the $\varepsilon \to +0$ limit.
The extrapolation procedure, as well as the 
special method for  integrals with quasi-singularities encountered when solving
Eqs.~\eqref{eq:U}, is described in detail in our previous works
\cite{deltuva:12c,carbonell:14a,deltuva:14b}.

The $pp$ Coulomb force is included using the method of screening and renormalization
\cite{alt:80a,deltuva:07b} where the screening radius $R = 12$ to 16 fm is found to be sufficient
to achieve convergence  for the Coulomb-distorted short-range part of the amplitude.
The obtained results are  well converged with respect to the partial-wave
expansion. When solving  Eqs.~\eqref{eq:U} we take into account isospin-singlet 
$2N$ partial waves
with  total angular momentum $j_x \leq 4$ and  isospin-triplet $2N$
partial waves  with orbital angular momentum $l_x \leq 7$, $3N$ partial
waves with spectator orbital angular momentum $l_y \leq 7$ and 
total angular momentum $J \leq \frac{13}{2}$, and $4N$ partial waves
with 1+3 and 2+2 orbital angular momentum $l_z \leq 8$. 
Initial and final deuteron-deuteron states with relative orbital angular momentum  
$L \leq 6$ are sufficient for the calculation of observables except
at the 25.3 MeV beam energy where we take into account also the states up to
$L \leq 8$ that yield a small but visible contribution.

\section{Results \label{sec:res}}

\begin{figure*}[!]
\begin{center}
\includegraphics[scale=0.68]{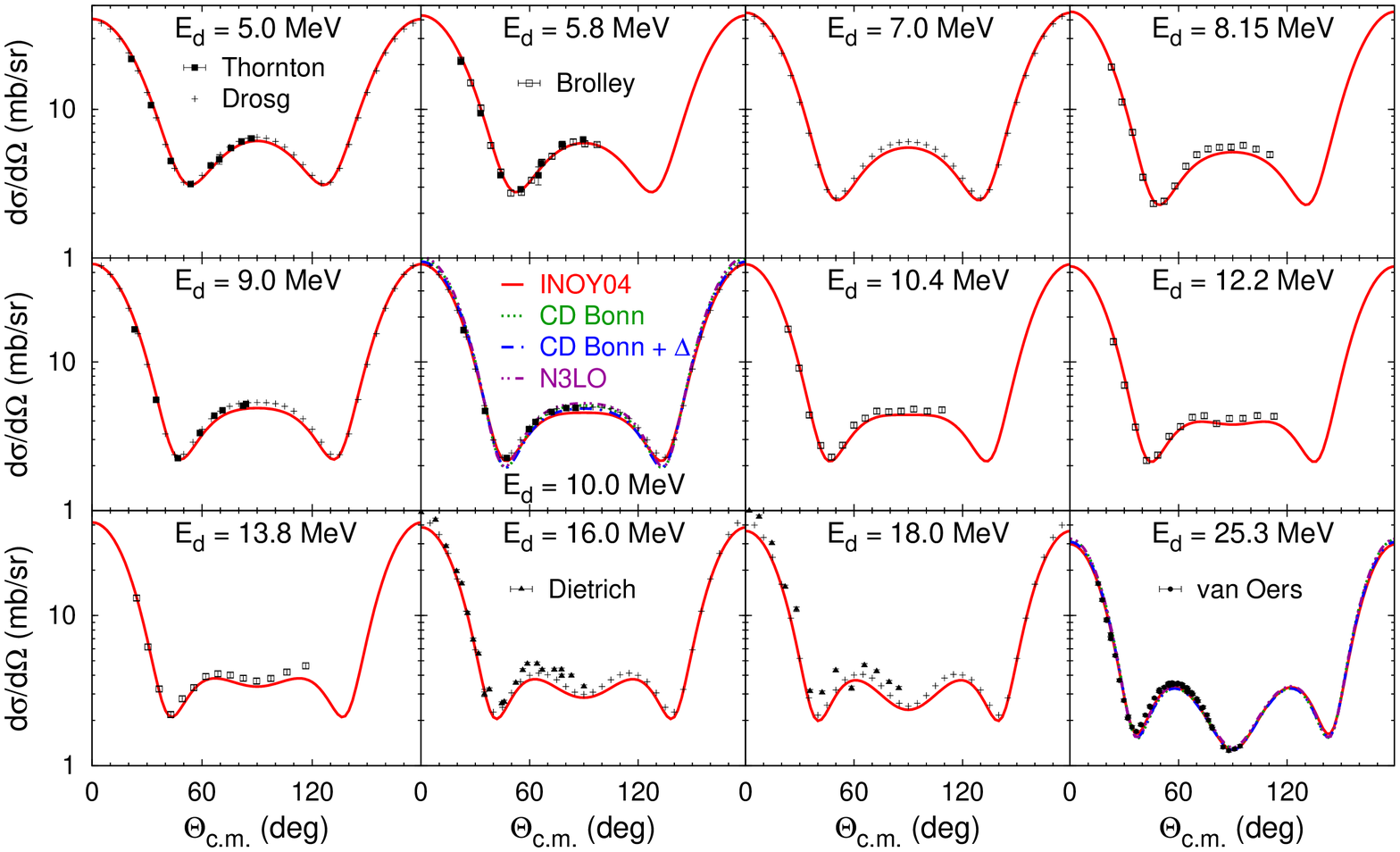}
\end{center}
\caption{\label{fig:sn} (Color online)
Differential cross section for the \Hdn reaction as a function
of the c.m. scattering angle at 
deuteron beam energies ranging from 5.0 to 25.3 MeV.
Curves are as in Fig.~\ref{fig:sp}.
 Experimental data  are from 
Refs.~\cite{thornton:69} ($\blacksquare$),  \cite{drosg:15} ($+$),
\cite{brolley:57} ($\Box$), \cite{dietrich:72} ($\blacktriangle$),
and \cite{vanoers:63b} ($\bullet$). }
\end{figure*}

The scattering of two deuterons, as compared to nucleon-trinucleon collisions,
 is more challenging from the computational point of view but also interesting physicswise.
Since deuterons are loosely bound and spatially large objects, 
their collision involves more partial waves and gives rise
to much higher breakup cross sections than encountered in other  $4N$
reactions initiated by nucleons and trinucleons.

The threshold for one (two) deuteron breakup corresponds to the deuteron beam energy 
$E_d = 4.45$ MeV (8.90 MeV).
We calculate differential cross sections $d\sigma/d\Omega$ and various spin observables
for ${}^2\mathrm{H}(d,p){}^3\mathrm{H}$ and  ${}^2\mathrm{H}(d,n){}^3\mathrm{He}$
reactions
as functions of the nucleon center-of-mass (c.m) scattering angle $\Theta_\cm$
 at  $E_d$ ranging from 5.0 to 25.3 MeV.  
At all considered energies the results are obtained using  the realistic
inside-nonlocal outside-Yukawa (INOY04) potential  of Doleschall
\cite{doleschall:04a,lazauskas:04a} since it  nearly
reproduces the experimental values of the $\He$  and $\Hh$
binding energies with no
additional $3N$ force. To investigate the dependence of the results on
the interaction model, at  $E_d = 10$ and 25.3 MeV, we also show the
predictions obtained with other high-precision $NN$ potentials.
These  are the  chiral effective field theory potential
at next-to-next-to-next-to-leading order (N3LO) \cite{entem:03a}, the
charge-dependent Bonn potential (CD Bonn)
\cite{machleidt:01a}, and its extension CD Bonn + $\Delta$
\cite{deltuva:03c} that explicitly includes the excitation of a nucleon to
a $\Delta$ isobar.  This mechanism generates  effective $3N$ and $4N$
forces that are  mutually consistent but quantitatively still
insufficient to reproduce $3N$ and $4N$ binding energies, although
they reduce the discrepancy \cite{deltuva:08a}.  The predictions
 for $\He$  and $\Hh$ binding energies for all employed force models
are collected in Table~\ref{tab:1}.

\begin{table}[htbp]
\begin{ruledtabular}
\begin{tabular}{l*{2}{c}}
 & $B(\Hh)$ & $B(\He)$    \\  \hline
N3LO        & 7.85 & 7.13 \\
CD Bonn     & 8.00 & 7.26  \\
CD Bonn + $\Delta$  & 8.28 & 7.53  \\
INOY04      & 8.49 & 7.73 \\
Experiment  & 8.48 & 7.72  
\end{tabular}
\end{ruledtabular}
\caption{$\Hh$ and $\He$ binding energies (in MeV)
for different $NN$ potentials.}
\label{tab:1}
\end{table}

\begin{figure*}[!]
\begin{center}
\includegraphics[scale=0.62]{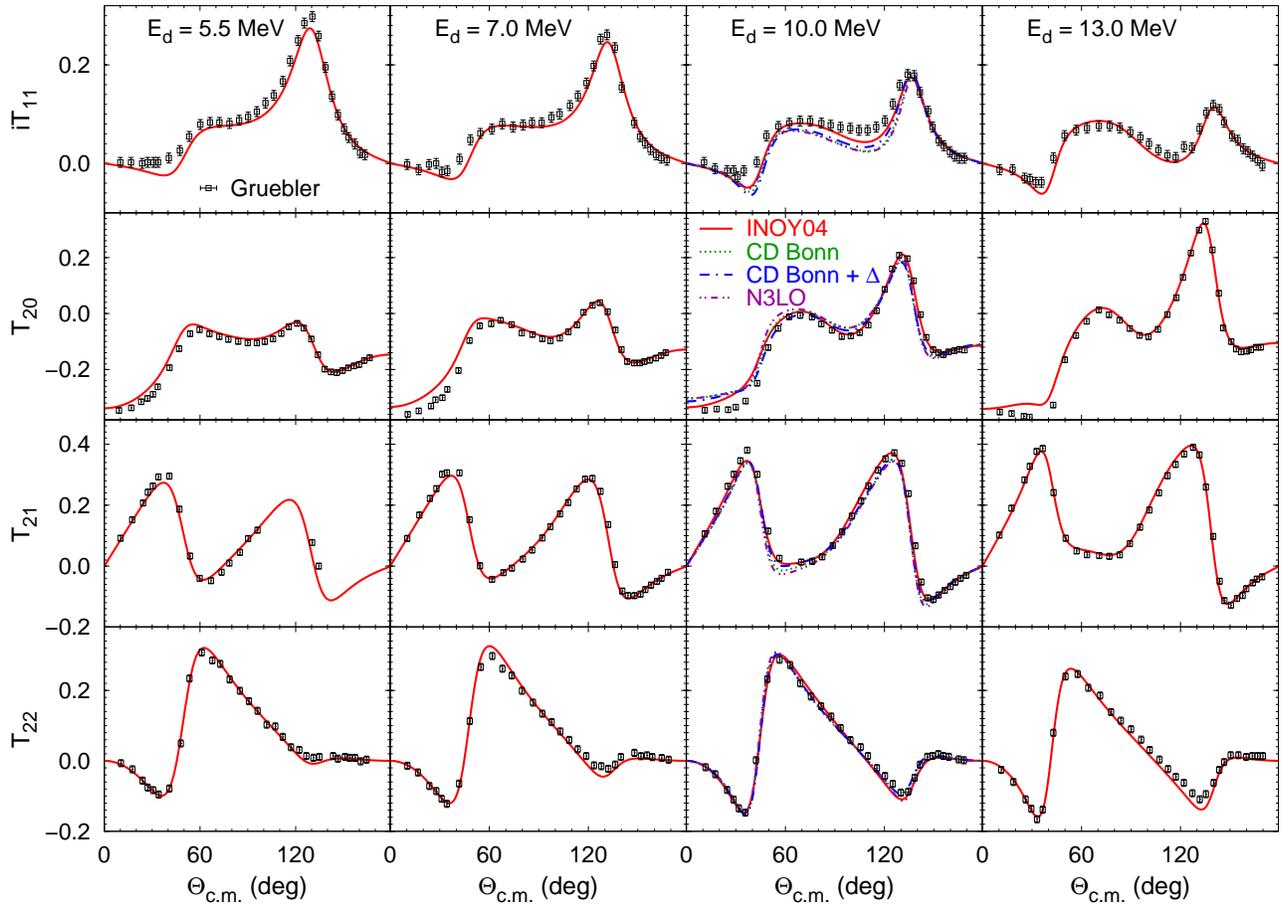}
\end{center}
\caption{\label{fig:tp} (Color online)
Deuteron analyzing powers for the \Hdp reaction
 at $E_d = 5.5$, 7.0, 10.0, and 13.0 MeV. 
Curves are as in Fig.~\ref{fig:sp}.
Experimental data are from  Ref.~\cite{gruebler:81a}.}
\end{figure*}

\begin{figure*}[!]
\begin{center}
\includegraphics[scale=0.62]{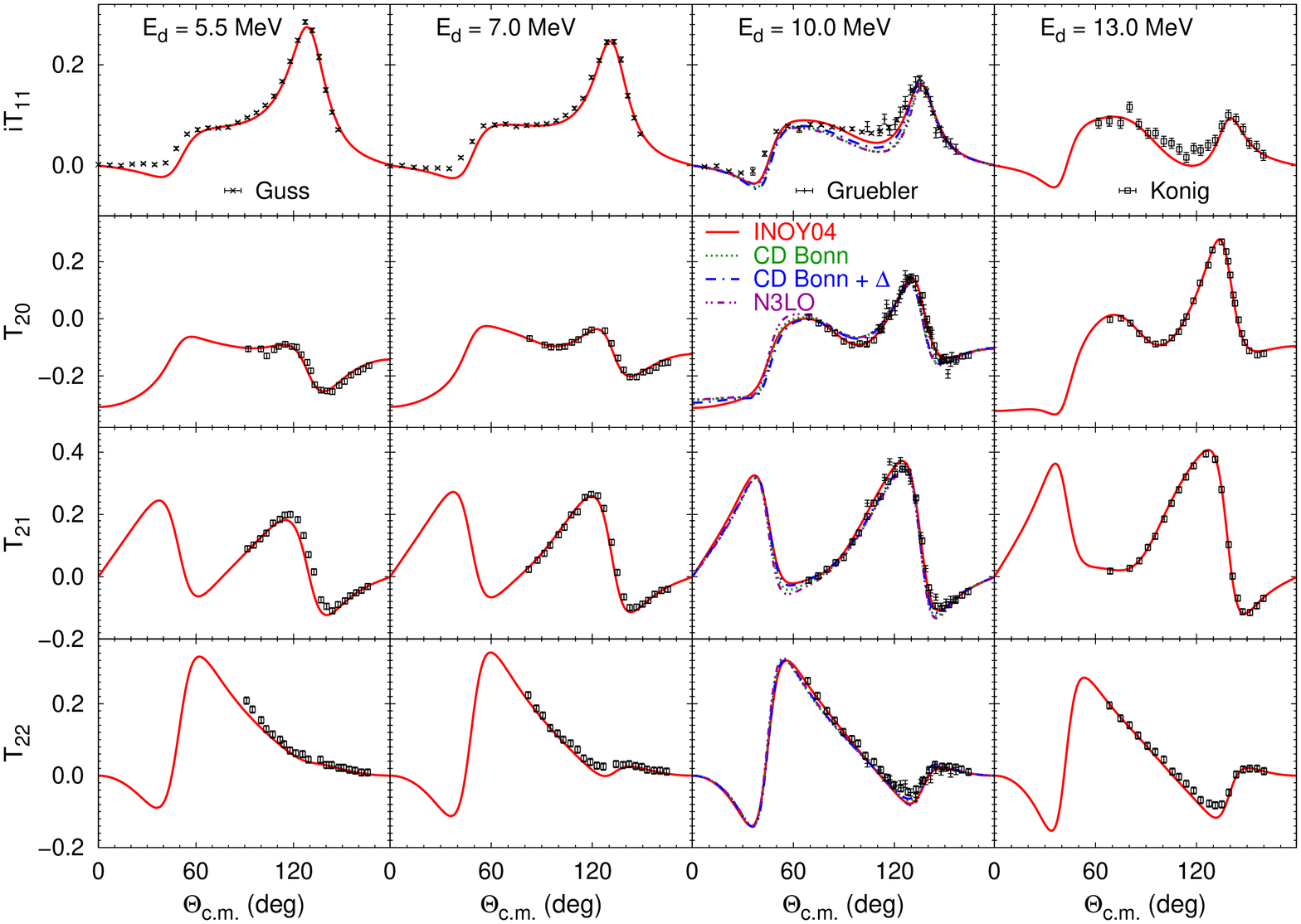}
\end{center}
\caption{\label{fig:tn} (Color online)
Deuteron analyzing powers for the \Hdn reaction
 at $E_d = 5.5$, 7.0, 10.0, and 13.0 MeV. 
Curves are as in Fig.~\ref{fig:sp}.
Experimental data are from  Refs.~\cite{guss:83} ($\times$),
\cite{gruebler:72a} (+), and \cite{konig:79} ($\Box$)}
\end{figure*}

\begin{figure}[!]
\begin{center}
\includegraphics[scale=0.57]{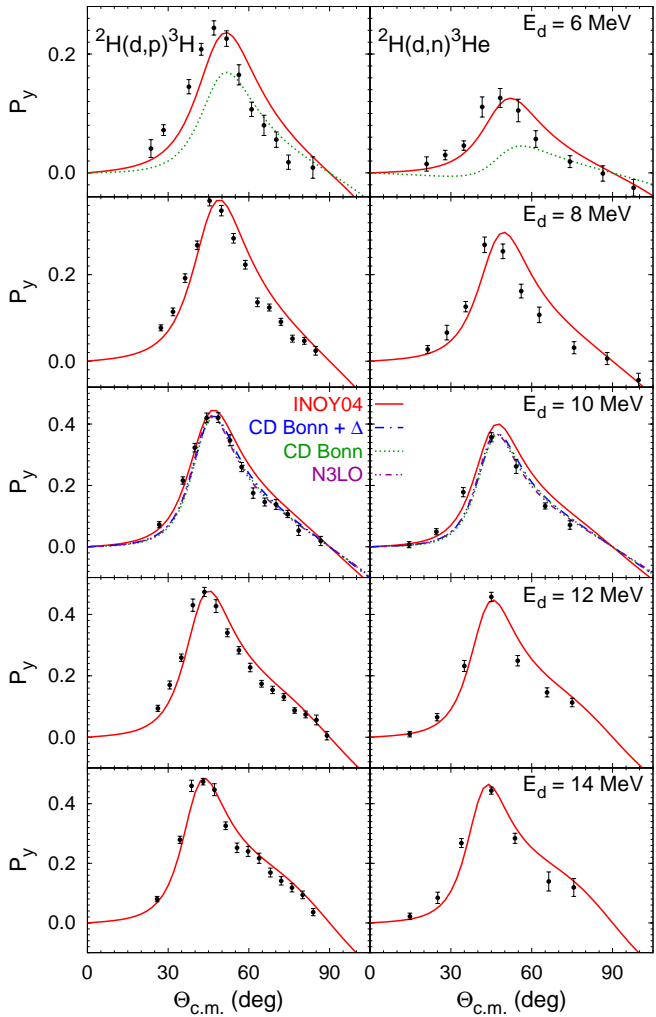}
\end{center}
\caption{\label{fig:Npol} 
Outgoing nucleon polarization 
of the \Hdp (left) and \Hdn (right)
transfer reactions at deuteron beam energies ranging from 6 to 14 MeV.
Curves are as in Fig.~\ref{fig:sp}. Experimental data are from 
Refs.~\cite{hardekopf:72b,spalek:72} for the \Hdp and \Hdn reactions, respectively.}
\end{figure}

\begin{figure}[!]
\begin{center}
\includegraphics[scale=0.6]{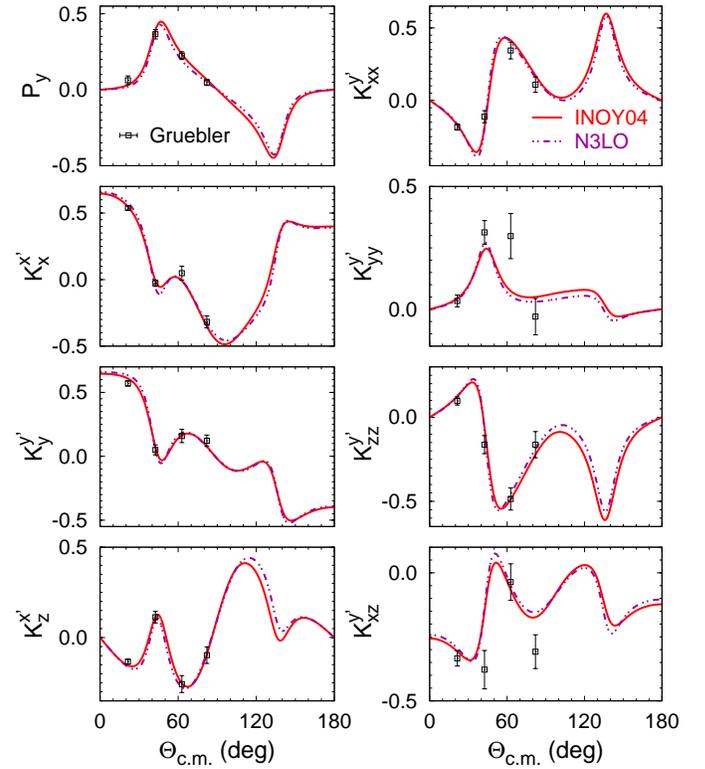}
\end{center}
\caption{\label{fig:pk} 
Outgoing proton polarization $P_y$ and
deuteron-to-proton polarization transfer coefficients 
for the \Hdp reaction at 10.15 MeV deuteron energy.
Curves are as in Fig.~\ref{fig:sp}.
Experimental data are from  Ref.~\cite{gruebler:74}.}
\end{figure}

\begin{figure}[!]
\begin{center}
\includegraphics[scale=0.6]{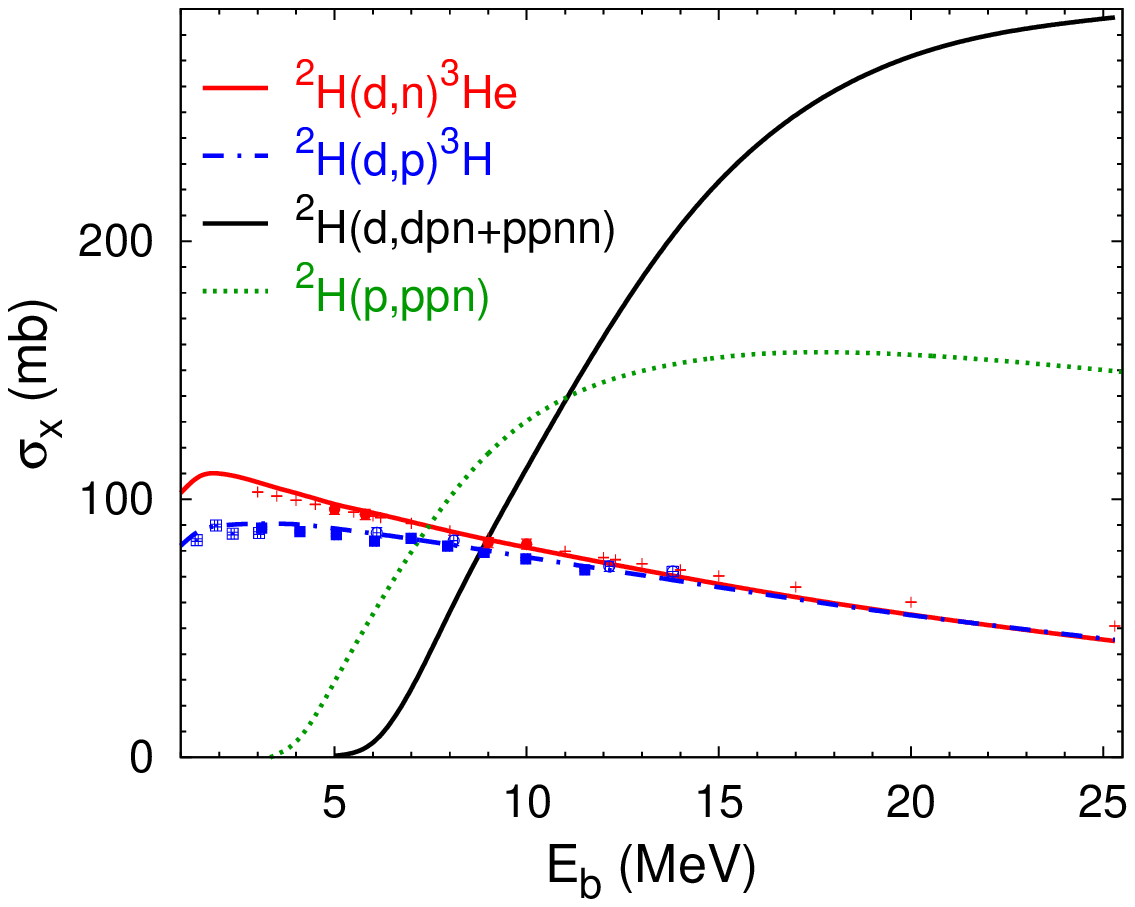}
\end{center}
\caption{\label{fig:tot} 
Total cross sections for transfer and breakup reactions in deuteron-deuteron
collisions as functions of the beam energy. In addition, 
results for the proton-deuteron breakup
are represented by the dotted curve.
Experimental data for transfer reactions are from  
Refs.~\cite{gruebler:72a,gruebler:81a,brolley:57,thornton:69,drosg:15}.}
\end{figure}

In Figs.~\ref{fig:sp} and  \ref{fig:sn} we show the differential cross sections for the transfer reactions  
${}^2\mathrm{H}(d,p){}^3\mathrm{H}$ and  ${}^2\mathrm{H}(d,n){}^3\mathrm{He}$, respectively. 
Due to identity of the two deuterons this
observable is symmetric with respect to $\Theta_\cm = 90^\circ$  and is peaked in the forward and
backward directions.
As the energy rises, the local maximum of $d\sigma/d\Omega$ at $\Theta_\cm = 90^\circ$  
evolves into a local minimum between two local maxima at about $60^\circ$ and $120^\circ$.
Over the whole energy and angular regime the INOY04 results closely follow the
experimental data \cite{gruebler:72a,gruebler:81a,brolley:57,vanoers:63b,thornton:69},
only slightly underpredicting some of them around $\Theta_\cm = 90^\circ$. 
The ${}^2\mathrm{H}(d,n){}^3\mathrm{He}$ data from Ref.~\cite{dietrich:72} are 
obviously wrong as they are inconsistent with the global evaluation \cite{drosg:15}.
The sensitivity to the $NN$ force model is visible at $E_d=10$ MeV, where it can reach 15\%,
but decreases with the energy, becoming insignificant at  $E_d=25.3$ MeV. As at very low energies
\cite{deltuva:07c}, the predictions roughly scale with the trinucleon binding energy.

The deuteron vector analyzing power $iT_{11}$ and
tensor analyzing powers $T_{20}$, $T_{21}$, and $T_{22}$
are shown in Figs.~\ref{fig:tp} and  \ref{fig:tn}  for the same transfer reactions
up to $E_d = 13$ MeV and compared with the experimental data
\cite{gruebler:81a,guss:83,gruebler:72a,konig:79};
 we are not aware of experimental data at higher energies. These observables
do not exhibit any symmetry with respect to $\Theta_\cm = 90^\circ$  as only the beam
deuteron is polarized.
We observe a few small discrepancies between predictions and data, mostly
for $iT_{11}$ and $T_{20}$ at $\Theta_\cm < 60^\circ$  in ${}^2\mathrm{H}(d,p){}^3\mathrm{H}$,
but the overall agreement with the data is impressive given the complicated angular dependence
 of the analyzing powers.
The sensitivity to the $NN$ force model is again visible, especially
for  $iT_{11}$, but shows no clear correlation with  the trinucleon binding energy.

The polarization $P_y$ of the outgoing nucleon for 
${}^2\mathrm{H}(d,p){}^3\mathrm{H}$ and  ${}^2\mathrm{H}(d,n){}^3\mathrm{He}$ reactions
is presented in Fig.~\ref{fig:Npol}. This observable is antisymmetric
with respect to $\Theta_\cm = 90^\circ$; hence we show only the angular regime
$\Theta_\cm < 105^\circ$ for $E_d$ up to 14 MeV
where the experimental data \cite{hardekopf:72b,spalek:72} 
are available. This observable exhibits a peak near $\Theta_\cm = 45^\circ$ 
whose height is well reproduced by the calculations with the INOY04 potential but the
position is shifted to larger angles by about 2 to 8 degrees; this shift is more pronounced
at lower beam energies. The sensitivity to the $NN$ potential 
at $E_d = 10$ MeV  is comparable to the one observed for analyzing powers,
but becomes significantly larger at lower energies. This is illustrated by the
CD Bonn predictions at $E_d = 6$ MeV and is consistent with the strong model dependence
found in Ref.~\cite{deltuva:10a} at $E_d = 3$ MeV.
We note that $P_y$ is equal to the nucleon analyzing power $A_y$ in the respective time reversed
reactions ${}^3\mathrm{H}(p,d){}^2\mathrm{H}$ and  ${}^3\mathrm{He}(n,d){}^2\mathrm{H}$
at the same energy $E$. Given that $A_y$ in low-energy nucleon-trinucleon scattering 
shows sizable discrepancies between predictions and datah for bot elastic and charge-exchange reactions
\cite{deltuva:15c}, some discrepancies for $A_y$ in the coupled nucleon transfer reactions
are expected. A similar conclusion can also be drawn regarding the
$NN$ force sensitivity in all coupled reactions. 

Measurement of double-polarization observables is highly complicated; thus, the
corresponding data are quite scarce. We are aware of only two experiments measuring
the angular dependence of deuteron-to-nucleon polarization transfer coefficients, i.e.,
$K_x^{x'}$, $K_x^{z'}$, $K_z^{x'}$, $K_z^{z'}$, $K_y^{y'}$, and $K_{yy}^{y'}$ 
in the ${}^2\mathrm{H}(\vec{d},\vec{n})\He$ reaction at $E_d = 10$ MeV 
\cite{salzman:74}, and $K_x^{x'}$, $K_z^{x'}$, $K_y^{y'}$, 
$K_{xx}^{y'}$,  $K_{yy}^{y'}$,  $K_{zz}^{y'}$ , and $K_{xz}^{y'}$ 
in the ${}^2\mathrm{H}(\vec{d},\vec{p})\Hh$ reaction at $E_d = 10.15$ MeV 
\cite{gruebler:74}. The former set is already analyzed in our earlier
work \cite{deltuva:15a}. Despite the very complex angular dependence of
deuteron-to-neutron polarization transfer coefficients having up to six local extrema,
a good  agreement between theory and experiment and little sensitivity to the $NN$ force
model are found. A similar situation holds also for the
 ${}^2\mathrm{H}(\vec{d},\vec{p})\Hh$ reaction at $E_d = 10.15$ MeV. 
We present our predictions using the INOY04 and N3LO potentials  in Fig.~\ref{fig:pk}. 
There are only four data points for each observable, 
and almost all of them lie on theoretical curves,
the exceptions being one or two points for  $K_{yy}^{y'}$ and $K_{xz}^{y'}$
having quite large error bars. Thus, one may conclude that the overall description of
polarization transfer data for both  ${}^2\mathrm{H}(\vec{d},\vec{n})\He$
and ${}^2\mathrm{H}(\vec{d},\vec{p})\Hh$ reactions
is quite satisfactory.

Finally, in Fig.~\ref{fig:tot}  we show the energy dependence of the total  cross section 
(integrated over the scattering angle) for both transfer reactions 
and the sum of the three- and four-cluster breakup. 
An explicit calculation of the breakup amplitudes is highly demanding, but the total
breakup cross section is obtained as the difference between the full
and all two-cluster channels cross sections, applying optical theorem
for amplitudes calculated with a sufficiently large Coulomb screening radius
\cite{deltuva:14b}. The comparison with data for
the transfer cross sections is reasonably good as could be expected 
from the comparison of differential cross sections in Figs.~\ref{fig:sp} and  \ref{fig:sn}.
We are not aware of total breakup cross section data. We predict a rapid
increase of the  breakup cross section at low energies, exceeding the
transfer cross sections around $E_d = 9$ MeV and becoming the most important
inelastic channel. For example, at $E_d = 25.3$ MeV the  breakup cross section
reaches 287 mb, more than 6 times larger than each of the transfer cross sections.
In the same figure we show the total proton-deuteron
breakup cross section as a function of the proton beam energy. This reaction has 
a lower threshold, but beyond 12 MeV the deuteron-deuteron breakup
cross section exceeds the one for the proton-deuteron breakup,
reaching a factor of 2 between them at 25 MeV. It is interesting to note that this factor of two 
at higher energies may be simply conjectured as resulting from the sum of projectile deuteron 
breakup and target deuteron breakup, given the symmetry of the deuterons in the entrance channel.

\section{Summary \label{sec:sum}}

In the present manuscript we show the results of our numerical
calculations aiming to describe the world data for both transfer
reactions, ${}^2\mathrm{H}(d,p){}^3\mathrm{H}$ and
${}^2\mathrm{H}(d,n){}^3\mathrm{He}$, in the energy region above the three-
and four-body breakup threshold, and present 
predictions for the total deuteron-deuteron breakup cross section up to $E_d = 25$
MeV. The calculations include realistic $NN$ force models and the
Coulomb interaction between protons.  Given the large size of the
deuteron, elastic scattering of two deuterons 
can be expected to be peripheral and 
display little sensitivity to the choice of $NN$ interaction. Nevertheless
this argumentation is not completely valid for the transfer reactions
where threshold positions depend on the $NN$ force. Furthermore,
due to the weak binding, deuteron-deuteron reactions are more demanding computationally as
they involve more partial waves and give rise to higher breakup cross sections. 
In contrast to nucleon-trinucleon reactions, given the symmetry of the two deuterons in the initial channel, 
only states that are symmetric under the exchange of the two deuterons contribute the observables 
which by itself curtails the number of combinations through which the $NN$ force affects the 
transfer observables. Another aspect in favor of deuteron-deuteron reactions relative to nucleon-trinucleon 
is that the threshold is well above the region where the four-nucleon system displays a complex structure of resonances.
Nevertheless, these features do not allow for a reasonable simplification of the AGS equations \eqref{eq:U}, 
as all the terms are important and cannot be neglected.

This being said one cannot avoid concluding that the results we obtain provide a very impressive description 
of the world data for the  ${}^2\mathrm{H}(d,p){}^3\mathrm{H}$ and
${}^2\mathrm{H}(d,n){}^3\mathrm{He}$ reactions up to $E_d = 25$ MeV. Given the complex structure of these data, 
not seen in nucleon-trinucleon or nucleon-deuteron scattering, discrepancies are very few and limited to a few observables 
such as $iT_{11}$ and $P_y$ and small angular regions, in spite of the large number of maxima and minima displayed by the data. 
Also double-polarization experiments are well described by the calculations. 

With this work we have completed the study of all possible two-cluster reactions in the four-nucleon system
using realistic pairwise potentials, and have brought the
four-nucleon scattering problem to the same degree of development that
three-nucleon scattering obtained already a number of
years ago. Further progress is still needed to calculate the breakup observables, but the lack of accurate 
breakup data makes the effort superfluous for now. In a recent article \cite{fonseca:17a} we reviewed 
our previous work and suggested  that the best theoretical laboratory to study $NN$ force models 
is in fact the low-energy region where there are a number of four-nucleon resonances in both  $T=0$ 
and $T=1$ total isospin configurations. Although four-nucleon calculations are more time consuming 
and pose greater numerical challenges, one finds a greater sensitivity to 
$NN$ force models  as compared to the three-nucleon system.

The authors thank M. Drosg for the discussion of experimental data.
A.D. acknowledges partial support from Lietuvos Mokslo Taryba (Research Council
of Lithuania) under Project. No. KEL-15018.


\end{document}